\documentclass[here,epsfig,referee]{eas}
\usepackage{graphicx}
\def\simlt{\mathrel{\rlap{\lower 3pt\hbox{$\sim$}}\raise 2.0pt\hbox{$<$}}}
\def\simgt{\mathrel{\rlap{\lower 3pt\hbox{$\sim$}} \raise 2.0pt\hbox{$>$}}}

\def\Msun{M_{\odot}}

\def\gtsima{$\; \buildrel > \over \sim \;$}
\def\ltsima{$\; \buildrel < \over \sim \;$}
\def\gtrsim{\lower.5ex\hbox{\gtsima}}
\def\lesssim{\lower.5ex\hbox{\ltsima}}
\def\url#1{{\ttfamily\def\/{/\discretionary{}{}{}}#1}}
\def\etal{{\it et al.~}}

\begin{document}

\title{Numerical Simulations of the Metallicity Distribution in Dwarf
  Spheroidal Galaxies}
\runningtitle{Metallicity Distribution of dSphs}
\author{E. Ripamonti}
\address{Kapteyn Astronomical Institute, University of Groningen;
  \email{ripa@astro.rug.nl}}
\author{E. Tolstoy}\sameaddress{1}
\author{A. Helmi}\sameaddress{1}
\author{G. Battaglia}\sameaddress{1}
\author{T. Abel}\address{Kavli Institute for Particle Astrophysics and
  Cosmology, Stanford Linear Accelerator Center}

\begin{abstract}
Recent observations show that the number of stars with very low
metallicities in the dwarf spheroidal satellites of the Milky Way is
low, despite the low average metallicities of stars in these systems. We
undertake numerical simulations of star formation and metal enrichment
of dwarf galaxies in order to verify whether this result can be
reproduced with ``standard'' assumptions. The answer is likely to be
negative, unless some selection bias against very low metallicity stars
is present in the observations.
\end{abstract}
\maketitle

\section{Introduction}

The study of the formation and evolution of galaxies is one of the most
important issues of present-day astronomy. The currently favoured models
suggest that large galaxies such as the Milky Way formed through the
hierarchical accretion of a number of smaller objects.

One possible way to test this scenario is to focus on dwarf galaxies: in
fact, it is reasonable to expect that their assembly was considerably
less complicated than that of the Milky Way, and easier to understand.

The nine dwarf spheroidal (dSph) satellites of the Milky Way are ideal
targets in this respect: theoretically, they might be the ``fossil''
remnants of the ``building blocks'' which ended up inside the Milky Way;
observationally, they can be studied in much greater detail than more
distant objects.

For example, it has been possible to study their stellar populations,
extracting informations about their star formation (SF) and chemical
evolution histories (see e.g. the review by Mateo 1998), which turned
out to be quite varied. However, all of them contain a population of
very old stars, and all of them exhibit low mean metallicities (Grebel
\& Gallagher 2004).

Large observational programs (such as DART, i.e. Dwarf Abundances and
Radial-velocities Team) are measuring, for the first time, the stellar
metallicity distribution in dSph galaxies (Tolstoy \etal 2004; Koch
\etal 2006; Battaglia \etal 2006). Despite the low average metallicities
(consistent with previous estimates), out of about 2000 stars which were
observed in four different galaxies (Carina, Fornax, Sculptor, and
Sextans), none of them turned out to have a metallicity lower than
[Fe/H]=-3, which is quite surprising (Helmi \etal 2006).

Here, we use numerical simulations of chemical enrichment of dwarf
galaxies in order to investigate whether this dearth of very metal poor
stars (VMPSs, i.e. stars with ${\rm [Fe/H]}\leq-3$) is consistent with
the simple hypothesis that the gas in the dwarf galaxies was completely
self-enriched in metals (i.e. that the gas metallicity when SF started
in these galaxies was essentially 0), and that the IMF of these galaxies
was always given by a Salpeter power law extending from 0.1 to 100
$\Msun$.

\section{The Simulations}

\subsection{The Method}
We modified the public SPH code Gadget (Springel \etal 2001;
Springel 2005) in order to include the treatment of gas cooling, SF,
supernova (SN) and stellar wind feedback, and metal enrichment of
the inter-stellar medium. A complete description will be given in
Ripamonti \etal 2006. Here it is sufficient to say that the gas cooling
rate was taken from Sutherland \& Dopita (1993) (if $T\geq10^4\;{\rm
K}$; otherwise it was assumed to be 0), the SF recipe
assumes a Schmidt law (see e.g. Thacker \& Couchman 2000), the stellar
lifetimes are taken from the Geneva evolutionary tracks (e.g. Schaller
\etal 1992), and that SN energies and yields are from Woosley \& Weaver
(1995).

\subsection{Sculptor}
Our simulations were aimed at reproducing the chemical properties of the
Sculptor dSph (hereafter, Scl), rather than those of the full sample of
Helmi \etal (2006). The reason for this choice is that the low
metallicity tail of Scl extends to slightly lower metallicities
than those of the other three dSphs (a fact which will strengthen our
conclusions). Furthermore, the SF history of Scl
appears to have lasted only a few Gyr, and after this initial
period it appears to have stopped, as no stellar population younger than
about 10 Gyr has been detected: such a simple history should be
relatively easy to reproduce.

The total mass of Scl was quoted to amount to a few $\times10^7\,\Msun$
(Queloz \etal 1995), but more recent estimates (Battaglia \etal 2006)
have put it at a much higher value ($\sim 5\times10^8\,\Msun$). Here we
report the results of simulations where its total mass was assumed to be
$M_{halo}=10^8\,\Msun$: such a value is low when compared to recent
measurements, but this should have only a small effect on the
metallicity distribution of the stars in the galaxy.

\begin{figure}
\includegraphics[height=10truecm,angle=270]{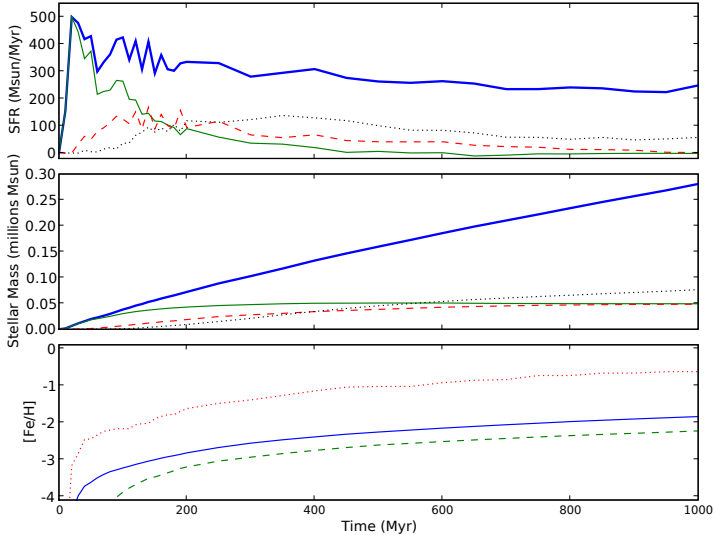}
\caption{Evolution of the stellar properties in one of our
  simulations. Top panel: total SF rate (solid thick), SF rate of stars
  with ${\rm [Fe/H]}\leq-3$ (thin solid), SF rate of stars with
  $-3.0<{\rm [Fe/H]}\leq-2.5$ (dashed), SF rate of stars with
  $-2.5<{\rm [Fe/H]}\leq-2$ (dotted). Central panel: total mass of stars,
  and mass of stars in metallicity ranges (symbols as in the top
  panel). Bottom panel: maximum (dotted), average (solid), and median
  (dashed) of stellar metallicities.}
\end{figure}

\subsection{Simulation setup and parameters}
At the beginning of our simulations we assume that all the baryons are
in gaseous form, and that both the gas and DM follow a NFW profile
(Navarro \etal 1997) with concentration $c=10$ and virial
radius $R_{200}=1.5\,{\rm kpc}$ (approximately coincident with the
present-day tidal radius of Scl). We place 20000 DM particles and 100000
gaseous particles within twice the virial radius. Since we assume that a
mass $M_{halo}$ is entirely enclosed within $R_{200}$, the total mass
included in each simulation is about $1.4 M_{halo}$, of which a fraction
$\Omega_{DM}/(\Omega_{DM}+\Omega_b)\simeq0.825$ is in the DM component,
and a fraction $1-\Omega_{DM}/(\Omega_{DM}+\Omega_b)\simeq0.175$ is in
the baryonic component ($\Omega_{DM}\simeq0.198$, and
$\Omega_b\simeq0.042$ are the cosmological density parameters of DM and
baryons; see Spergel \etal 2006).

The initial velocities were assigned according to the recipe for a
spherical halo described in Hernquist (1993), and the gas particles were
assumed to be cold.

The main parameters of our simulations were related to SF and
feedback. They include the typical mass of stellar particles
($50\Msun$), the SF efficiency $C_{sfr}$ (which was varied in the range
$0.001-0.1$; see e.g. Thacker \& Couchman 2000), the typical energy of a
SN explosion which is transferred to neighbouring gas particles~($10^{50}$~erg)\footnote{In experiments conducted with a SN energy of $10^{51}$
erg, the feedback completely stopped the SF after much less than one
Gyr, preventing the formation of more than a few $10^4\,\Msun$ of
stars.}, and the fraction $F_{ret}$ of the metals ejected in a SN
explosion which is retained by the galaxy (we tested $F_{ret}=1$ and
$F_{ret}=0.1$; this second value is justified by the results of Mac Low
\& Ferrara 1999, which found that metals from the SN ejecta can
escape from the galaxy far more easily than the rest of the gas).

\begin{figure}
\includegraphics[height=10truecm,angle=270]{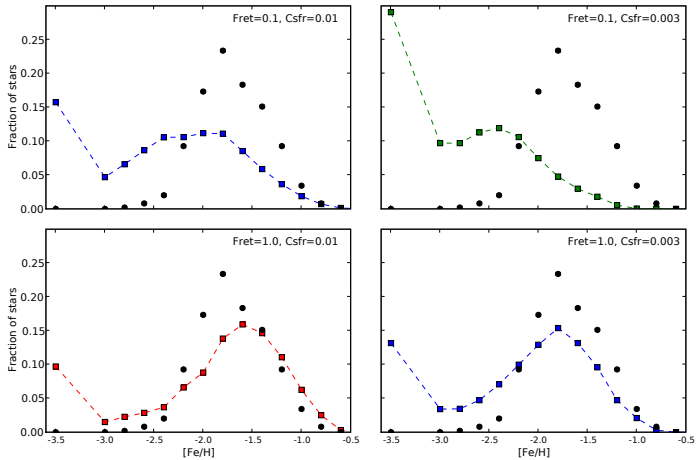}
\caption{Metallicity distribution after 1 Gyr for stars in four
  different simulations (squares connected by dashed lines), differing
  for the values of $F_{ret}$ and $C_{sfr}$ indicated in each panel. The
  dots (whose size is comparable to the error bars) show the metallicity
  distribution of 496 stars in Scl. The leftmost point (at
  ${\rm [Fe/H]}=-3.5$) actually groups together all the VMPSs. Values are
  normalized to the total number of stars.}
\end{figure}

\section{Results}
We ran a grid of simulations with different combinations of the above
parameters; each one was run for just 1 Gyr, because in all of them we
found that the formation of very low metallicity stars had essentially
stopped before that time (see Fig. 1). The mass of the stellar
component was always much smaller (typically, by a factor 3-10) than the
stellar mass in Scl, but SF in the simulated galaxy was
still active, even if only for stars with ${\rm [Fe/H]}\gtrsim-2.3$.

This fact must be kept into account when comparing the observed Scl
metallicity distribution with those produced by the simulations, because
the average stellar metallicity from the simulations is still
growing. In Fig. 2 we show such a comparison in four typical cases.

It is apparent that the fraction of VMPSs is always very high; it is
higher in models where a low value of $F_{ret}$ and an high value of
$C_{sfr}$ are assumed (which is unsurprising because such assumptions
correspond to a longer timescale for the metal enrichment of the
gas). In all the cases the fraction of VMPSs is difficult to reconcile
with the observations, even when a ``dilution'' by a factor 3-10 (due to
the future formation of a large number of stars) is
introduced. Furthermore, the models with $F_{ret}=1$, where this
discrepancy is lower, suffer from another problem at the high
metallicity end, since they produce an average metallicity which is too
high, at least if $C_{sfr}\geq0.01$.

In Fig. 3 we try to limit the effects of the unknown SF
after the first Gyr of evolution by looking just at the metallicity
distribution of stars with ${\rm [Fe/H]}\leq-2.3$, because in such metallicity
range SF is essentially complete by the time the simulations
are stopped. Here the excess of VMPSs appears less dramatic; but this is
mostly an artifact of the large error bars due to the low number (23) of
observed stars in this metallicity range. Furthermore, the {\it shape}
of the distributions appear to be different: the models fail to
reproduce the observed increase in the number of stars at
${\rm [Fe/H]}\gtrsim-2.5$, and predict a very large number of essentially
metal-free stars (when $F_{ret}=1$ is assumed, most of the VMPSs have
${\rm [Fe/H]}\leq-4$).

\begin{figure}
\includegraphics[height=10truecm,angle=270]{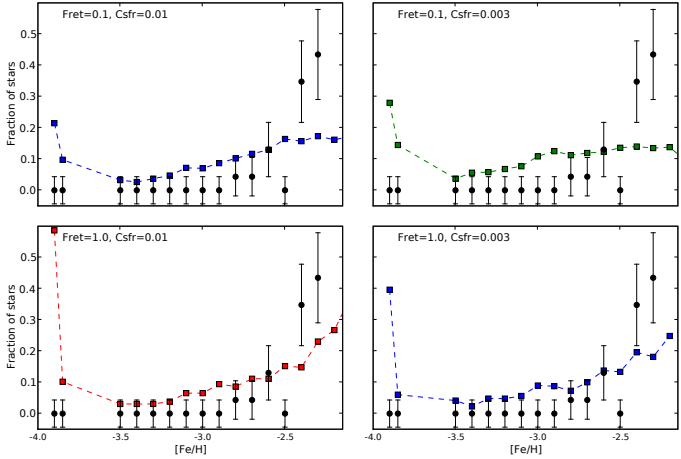}
\caption{Metallicity distribution for low metallicity stars after 1 Gyr
  in four different simulations (squares connected by dashed lines),
  differing for the values of $F_{ret}$ and $C_{sfr}$ indicated in each
  panel. Dots with error bars show the metallicity distribution of stars
  in Scl. The two leftmost bins refer to stars with ${\rm [Fe/H]}\leq-4$, and
  with $-4<{\rm [Fe/H]}\leq-3.5$. All the values are normalized to the number
  of stars with $-3\leq[Fe]/H\leq-2.3$.}
\end{figure}

\section{Discussion and conclusions}
The metallicity distribution obtained in our simulations is quite
different from the predictions of Lanfranchi \& Matteucci (2004), as
they predict a quite sharp drop at low metallicity (${\rm
[Fe/H]}\lesssim-2.5$), in agreement with observations (however, there is
significant disagreement at ${\rm [Fe/H]}\gtrsim-1.5$). This is probably
due to their assumptions about the ``infall'' history of gas inside the
galaxy (which implies a very low SF rate at early times, when the VMPSs
should form), and about the complete mixing of gas (so that there is no
spread in the age-metallicity relation). Instead, we have a large spread
(of the order of 1 dex) in the metallicities of stars which form after
the very early stages of our simulations; furthermore, we do not need to
assume an infall history for the gas, even if it can be argued that our
initial conditions are not completely realistic because of the
assumption that no star ever formed before the halo density profile
reached a NFW shape.

Our simulations indicate that the dearth of observed VMPSs in dSphs is
problematic. Apart from the hypothesis that observations are biased in
some unidentified way against the detection of VMPSs, possible solutions
might involve a pre-enrichment of the gas up to the ${\rm [Fe/H]}\sim-3$
level (see e.g. Helmi \etal 2006), or a difference between the present
and the primordial (metal-free) IMF, such as a suppression of the SF rate of
stars below $1\,\Msun$ in environments of very low metallicity (see
e.g. Omukai \etal 2005).


\end{document}